\documentclass[10pt,conference]{IEEEtran}
\usepackage{epsfig}
\usepackage{amsmath}
\usepackage{amsfonts}
\usepackage{amssymb}
\usepackage[all]{xy}
\usepackage{xcolor}
\usepackage{comment}
\usepackage{perso}
\usepackage{graphicx}
 \DeclareGraphicsExtensions{.eps}

\newtheorem{theo}{Theorem}

\newtheorem{rema}[theo]{Remark}

\newtheorem{exam}[theo]{Example}
\newtheorem{defi}[theo]{Definition}

\IEEEoverridecommandlockouts
\title{Binary Linear-Time Erasure Decoding for Non-Binary LDPC codes}
\author{Valentin~Savin, CEA-LETI, MINATEC, Grenoble, France, valentin.savin@cea.fr%
\thanks{This work has been partially supported by the French ANR grant
N° 2006 TCOM 019 (CAPRI-FEC project)}}
\date{}
\begin{document}
\maketitle

\begin{abstract}
In this paper, we first introduce the {\em extended binary
representation} of non-binary codes, which corresponds to a
cover\-ing graph of the bipartite graph associated with the
non-binary code. Then we show that non-binary codewords correspond
to binary codewords of the extended representation that further
satisfy some simplex-constraint: that is, bits lying over the same
symbol-node of the non-binary graph must form a codeword of a
simplex code. Applied to the binary erasure channel (BEC), this
description leads to a binary erasure decoding algorithm of
non-binary LDPC codes, whose complexity depends linearly on the
cardinality of the alphabet. We also give insights into the
structure of stopping sets for non-binary LDPC codes, and discuss
several aspects related to upper-layer FEC applications.
\end{abstract}


\section{Introduction}

Data loss recovery -- for instance, for content distribution
applications or for distributed storage systems -- is widely
addressed using erasure codes that operate at the transport/link or
the application layer of the communication system. Source data
packets are extended with repair packets that are used to recover
the lost data at the receiver. In this context, {\em Maximum
Distance Separable} (MDS) codes are ideal codes, in the sense that
decoding is possible as soon as the number of received packets
equals the number of source data packets. However, for large block
lengths, their decoding becomes untractable, and thus iteratively
decoded graph-based codes constitute the main alternative. Binary
{\em Low-Density Parity-Check} (LDPC) codes \cite{gall_phd}, with
iterative decoding, have been proven to perform asymptotically close
to the channel capacity \cite{richardson_design_2001}
\cite{luby2001eec}, while the decoding complexity per decoded bit is
independent of the code length. Tanner represented
 LDPC codes by sparse bipartite graphs, and showed that they
  can be generalized by replacing single parity check-nodes with
more general constraint-nodes \cite{Tann}. Nowadays, these codes are
referred as GLDPC codes and were recently investigated for the BEC
\cite{paolini:agl}, \cite{mila-foss-gldpc}.  Another class of
graph-codes, which have the attractive property of being able to
generate an infinite sequence of repair packets, are the {\em
rateless codes} proposed in \cite{luby2002lc}
\cite{shokrollahi2006rc}.  Over the past few years there also has
been an increased interest in
 non-binary LDPC codes due to their enhanced correction capacity.
 They were mainly investigated for physical-layer channels, but at this time only few works are dealing with the BEC
 \cite{rathi:det}, \cite{rathi:cen}, \cite{savin2008nbl}.
 Despite their performance, non-binary LDPC codes still have to
 overcome the obstacle of decoding complexity
 in order to become
 attractive for practical systems.

In this paper, we introduce the {\em extended binary representation}
of non-binary codes. From a graph point of view, the extended
representation corresponds to a covering graph of the bipartite
graph representing the non-binary code.
The covering graph represents a binary code, and we show that any
non-binary codeword can be lifted to a binary codeword of the
covering graph. This gives a one-to-one correspondence between
non-binary codewords and binary codewords of the covering graph that
are further constrained by a simplex code\footnote{A simplex code is
the dual of a Hamming code.} (that is, bits lying over the same
symbol-node of the non-binary graph must form a codeword of a
simplex code). By using the extended representation, we derive a
binary erasure decoding for the BEC, whose complexity depends
linearly on the cardinality of the alphabet, and which recover the
values of the erased bits from messages received from both simplex
and parity check constraints.


The paper is organized as follows. In section \ref{non_binary_ldpc}
we fix the notation used throughout the paper, and we review the
construction of non-binary LDPC codes and their decoding over the
BEC. The extended binary representation of non-binary codes is
introduced in section \ref{extended_representation}. In section
\ref{linear_time_decoding} we derive the binary erasure decoding of
non-binary LDPC codes, and we discuss stopping sets and several
aspects related to upper-layer FEC applications. Finally, section
\ref{conclusions} concludes the paper.

\section{Non-binary LDPC codes} \label{non_binary_ldpc}
We consider non-binary codes defined over an alphabet ${\cal A}$
with $q$ elements, where $q = 2^p$ is a power of $2$ (the last
condition is only assumed for practical reasons). We assume that
${\cal A}$ is endowed with a vector space structure over $\gf_2$
(the field with 2 elements), and we fix once for all an isomorphism
of vector spaces:
\begin{equation}
    \label{identify}
     {\cal A} \stackrel{\sim}{\longrightarrow} \gf_2^p
\end{equation}
Elements of ${\cal A}$ will also be called {\em symbols}, and we say
that $(x_0,\dots,x_{p-1})\in\gf_2^p$ is the {\em binary image} of
the symbol $X\in{\cal A}$ if they correspond to each other by the
above isomorphism.

Let $\mathbb{L} = \mcl{L}_{\gf_2}({\cal A})$ denote the algebra of
$\gf_2$-endomorphisms of ${\cal A}$. By evaluating elements of
$\mathbb{L}$ on symbols of ${\cal A}$ we get a left action of
$\mathbb{L}$ on ${\cal A}$, which will be denoted multiplicatively:
\begin{equation}
  \label{action}
    \mathbb{L} \times {\cal A} \rightarrow  {\cal A}:\ \  (h, X) \mapsto
    hX := h(X)
\end{equation}
Any matrix $H \in \mbf{M}_{M,N}(\mathbb{L})$ defines a
 code ${\cal C} \subset{\cal A}^N $:
\begin{eqnarray}
    \label{code_def}
  \mcl{C} \hspace{-2mm}&=\hspace{-2mm}& \ker(H)  \subset{\cal A}^N \\
     &=\hspace{-2mm}& \{ (X_1,\dots,X_N)\mid \sum_{n=1}^N h_{m,n}X_n = 0,\ \forall m = 1,\dots,M
     \}\nonumber
\end{eqnarray}

\begin{rema}\label{codes_gfq}
  Codes defined
  over $\gf_q$ -- the finite field with $q$ elements -- are
  a particular case of the above definition. The alphabet
  of these codes is ${\cal A} = \gf_q$, whose $\gf_2$-vector space structure  is
  inherited from the additive operation on $\gf_q$. Also, the internal field
  multiplication gives an embedding of $\gf_q$ as a vector subspace of $\mathbb{L} = \mcl{L}_{\gf_2}({\cal
  A})$. We say that {\em the code ${\cal C}$ is defined over $\gf_q$} if ${\cal C}$
  is defined as the kernel of a matrix $H\in \mbf{M}_{M,N}(\gf_q) \subset
  \mbf{M}_{M,N}(\mathbb{L})$. In this case ${\cal C}$ is a $\gf_q$-vector subspace of $\gf_q^N$.
\end{rema}

\subsection{The binary image of a non binary code}
A sequence of symbols $(X_1,\dots,X_N)\in{\cal A}^N$ may be mapped
into a binary sequence of length $Np$ via the isomorphism of
(\ref{identify}); this binary sequence will be referred as the
binary image of the given sequence of symbols. The binary images of
the codewords $(X_1,\dots,X_N)\in\mcl{C}$ form a linear binary code
$\mcl{C}_\txtscript{bin} \subseteq \gf_2^{Np}$, called the {\em
binary image of $\mcl{C}$}. The isomorphism of (\ref{identify}) can
also be used to further identify:
\begin{equation}
  \mbb{L} = \mcl{L}_{\gf_2}({\cal A}) \stackrel{\sim}{\rightarrow} \mcl{L}_{\gf_2}(\gf_2^p) = \mbf{M}_{p}(\gf_2)
\end{equation}
Thus, by replacing each entry of $H\in\mbf{M}_{M,N}(\mbb{L})$ with
its image under the above identification, we obtain a binary matrix
$H_\txtscript{bin}\in\mbf{M}_{Mp,Np}(\gf_2)$, which is the parity
check matrix of the binary code $\mcl{C}_\txtscript{bin}$.
\begin{rema} \label{bin_image_h}
  To avoid confusion, vectors will always be left-multiplied by a given
  matrix (unless the contrary is explicitly stated). Thus, if $h\in\mbb{L}$ and $m_h\in\mbf{M}_p(\gf_2)$ is its binary
  image, we have $hX = Y\Leftrightarrow m_h(x_0,\dots,x_{p-1})^{t} =
  (y_0,\dots,y_{p-1})^{t}$, for all $X,Y\in{\cal A}$.
\end{rema}

\subsection{Graphical representation}
The bipartite graph associated with a non-binary code ${\cal C}$,
denoted by ${\cal H}$, consists of $N$ {\em symbol-nodes} and $M$
{\em constraint-nodes}\footnote{These nodes are generally called
{\em check-nodes}. However, we will use {\em constraint-nodes} for
non-binary codes, and {\em check-nodes} for binary codes.}
representing respectively the $N$ columns and the $M$ rows of the
matrix $H$. A symbol-node and a constraint-node are connected by an
edge of ${\cal H}$ if the corresponding entry of matrix $H$ is a
non-zero element of $\mathbb{L}$ (note that the corresponding entry
is not assumed invertible!). Each edge of the graph is further
labeled by the corresponding non-zero entry of $H$.
We also denote by ${\cal H}(n)$ the set of constraint-nodes
connected to a given symbol-node $n \in \{ 1, 2, \dots, N \}$, and
by ${\cal H}(m)$ the set of symbol-nodes connected to a given
constraint-node $m \in \{ 1, 2, \dots, M \}$.

\subsection{Decoding over the BEC}
\label{gf_dec_bec} In this section we assume that a non-binary LDPC
code is used over the BEC($\epsilon$) -- the binary erasure channel
with erasure probability $\epsilon$. Thus, the length $N$ sequence
of encoded symbols is mapped into its binary image of length $Np$,
which is transmitted over the BEC; each bit from the binary image
being erased with probability $\epsilon$.

At the receiver part, the received bits are used to reconstruct the
corresponding symbols of the transmitted codeword.
Let $n$ be a symbol-node of the Tanner graph. We say that a symbol
$X\in{\cal A}$ is {\em eligible} for the node $n$, if the
probability of the $n^\txtscript{th}$ transmitted symbol being $X$
is non-zero. Tacking into consideration the channel output, the {\em
set of eligible symbols}, denoted by $\msr{E}_n$, consists of the
symbols whose binary images fit with the received bits (if any) of
the $n^\txtscript{th}$ transmitted symbol.
%
%
 These sets constitute the {\em a priori information} of the decoder. They are iteratively updated by exchanging
 messages between symbol and constraint-nodes in the graph. Each message is a subset of ${\cal A}$,
 representing a set of eligible symbols, either from the constraint-node or from the
 symbol-node perspective:
\begin{itemize}
  \item Each constraint-node $m$ represents a linear combination of symbol-nodes
  $n\in{\cal H}(m)$, whose coefficients are given by the corresponding edge
  labels. The constraint-node $m$ is verified if this linear
  combination is equal to zero. Therefore, for each $n\in {\cal
  H}(m)$ we can derive a set
  of eligible symbols, denoted by ${\cal E}_{m,n}$, according to the
  sets of eligible symbols ${\cal E}_{n'}$, with $n'\in {\cal H}(m)
  \setminus\{n\}$.
  \item On the other hand, each symbol-node $n$ is involved in several linear constraints given
  by the
  nodes $m\in{\cal
  H}(n)$, all of which must be verified. Therefore, we can update the set ${\cal E}_n$, by tacking into account the sets of eligible symbols
  ${\cal E}_{m,n}$, with $m\in {\cal H}(n)$.
\end{itemize}

Using the above notation, the iterative decoding for the BEC can be
expressed as follows (see also \cite{savin2008nbl}):
\begin{itemize}
\item {\bf constraint-node processing}

${\cal E}_{m,n} = \displaystyle\sum_{n'\in\mcl{H}(m)\setminus\{n\}}
h_{m,n'} {\cal E}_{n'}$

\bigskip
\item {\bf symbol-node processing}

${\cal E}_{n} = \displaystyle{\cal E}_n
\cap\left(\bigcap_{m\in\mcl{H}(n)} h_{m,n}^{-1} {\cal
E}_{m,n}\right)$

where $h_{m,n}^{-1} {\cal E}_{m,n} := \{ X\in {\cal A} \mid
h_{m,n}X\in {\cal E}_{m,n}\}$ (recall that $h_{m,n}$ is not assumed
 to be invertible).
\end{itemize}
These two steps are iterated as long as the cardinality of any
${\cal E}_n$ can be decreased. The decoding succeeds whenever all
the sets of eligible symbols ${\cal E}_{n}$ get cardinality $1$.
%
It can be seen that any set of eligible symbols, ${\cal E}_n$ or
${\cal E}_{m,n}$, is a $\gf_{2}$-affine subspace of ${\cal A}$; in
particular, its cardinal is a power of $2$.
\begin{rema}
  In the above description of the erasure decoding, a symbol-node $n$ send the same
  message ${\cal E}_n$ to all its neighbor constraint-nodes, violating the
  {\em extrinsic information principle} of a message-passing iterative
  decoding. However, the erasure decoding would not be changed by processing symbol-nodes in an extrinsic
  manner. This is due to the specificity of the BEC, which either erases a bit or
  transmits it correctly.
\end{rema}

\section{Extended binary representation of a non-binary LDPC code}
\label{extended_representation}

Let $\mathbb{Z}_q = \{0, 1,\dots, q-1\}$ denote the set of integers
modulo $q$. The bitwise XOR operation endows $\mathbb{Z}_q$ with a
vector space structure over $\mathbb{F}_2$, and the mapping
$\mathbb{Z}_q \rightarrow \gf_2^p$ that sends an integer into its
binary decomposition\footnote{We assume that the first bit of the
binary decomposition is the least significant bit}
 defines a vector space isomorphism.

Let $h\in \mathbb{L}$ and let $m_h \in \mbf{M}_{p}(\gf_2)$ be its
binary image. By using the above isomorphism, we obtain the
following endomorphism of $\mathbb{Z}_q$:
$$\Phi_h:\mathbb{Z}_q \stackrel{\sim}{\rightarrow} \gf_2^p \stackrel{^{t}\!m_h}{\longrightarrow} \gf_2^p \stackrel{\sim}{\rightarrow} \mathbb{Z}_q$$
where $^{t}\!m_h$ is the transpose of the matrix $m_h$. Thus,
$\Phi_h$
satisfies $\Phi_h(i\wedge j) = \Phi_h(i) \wedge \Phi_h(j)$, where
$\wedge$ is the bitwise XOR operation. The matrix $M_h\in
\mbf{M}_{q-1}(\gf_2)$ defined by:
$$M_h(i,j) = \left\{ \begin{array}{l}
  1, \mbox{ if } j = \Phi_h(i)\\
  0, \mbox{ otherwise}
  \end{array}\right.$$
  where $(i,j) \in \mathbb{Z}_q^{*} \times \mathbb{Z}_q^{*}$, is called {\em the extended matrix representation}
  of $h$.
When $h$ is an invertible element of ${\mathbb L}$ (or,
equivalently, $m_h$ is an invertible matrix of
$\mbf{M}_{p}(\gf_2)$), $\Phi_h$ induces a permutation of ${\mathbb
Z}_q^{*}$, thus $M_h$ is a permutation matrix.
\begin{rema}
  The use of $\mathbb{Z}_q$ in the above
  definition is only intended for indexing rows and
  columns of $M_h$ by integers rather than by symbols of ${\cal A}$ or by
  elements of $\gf_2^p$.
\end{rema}
\begin{exam}
  Assume that $p=3$, and let $h\in{\mathbb L}$ with binary image $m_h$ given by:
  $$m_h  = \left( \begin{array}{ccc}
           1 & 0 & 1 \\
           1 & 1 & 1 \\
           0 & 1 & 1
           \end{array} \right)$$
  The rows of $m_h$ define respectively $\Phi_h(1)$,
  $\Phi_h(2)$, and $\Phi_h(4)$. Thus, $\Phi_h(1) = 5$ is the integer whose
  binary decomposition is given by the first row of $m_h$, and similarly $\Phi_h(2) = 7$ and $\Phi_h(4)
  = 6$. Finally:
  \begin{itemize}
  \item $\Phi_h(3) = \Phi_h(1) \wedge \Phi_h(2) = 2$
  \item $\Phi_h(5) = \Phi_h(1) \wedge \Phi_h(4) = 3$
  \item $\Phi_h(6) = \Phi_h(2) \wedge \Phi_h(4) = 1$
  \item $\Phi_h(7) = \Phi_h(1) \wedge \Phi_h(2) \wedge \Phi_h(4) = 4$
  \end{itemize}
\end{exam}

Before defining the extended binary representation a non-binary
code, let us further develop this example. Consider now a non-binary
code defined by a single linear constraint:
$$h_1X + h_2Y +h_3Z = 0,$$
where $h_1,h_2,h_3\in {\mathbb L}$, and $X, Y, Z\in {\cal A}$.
Assume that after replacing $h_1,h_2,h_3$, and $X, Y, Z$ by their
binary images, the above equation becomes (see also Remark
\ref{bin_image_h}):
$$\left( \begin{array}{*{3}{@{\;\!}c@{\;\!}}}
         1 & 0 & 1 \\
         1 & 1 & 1 \\
         0 & 1 & 1
  \end{array} \right)
  \left( \begin{array}{*{1}{@{\;\!}c@{\;\!}}}
         x_0 \\ x_1 \\ x_2
  \end{array} \right) +
  \left( \begin{array}{*{3}{@{\;\!}c@{\;\!}}}
         0 & 1 & 0 \\
         0 & 1 & 1 \\
         1 & 0 & 1
  \end{array} \right)
  \left( \begin{array}{*{1}{@{\;\!}c@{\;\!}}}
         y_0 \\ y_1 \\ y_2
  \end{array} \right) +
  \left( \begin{array}{*{3}{@{\;\!}c@{\;\!}}}
         0 & 1 & 1 \\
         1 & 1 & 0 \\
         1 & 1 & 1
  \end{array} \right)
  \left( \begin{array}{*{1}{@{\;\!}c@{\;\!}}}
         z_0 \\ z_1 \\ z_2
  \end{array} \right)
  = 0$$
or equivalently:
\begin{equation}\begin{array}{*{6}{@{\;\!}c@{\;\!}}}
  (x_0+x_2)         & + &  y_1      & + & (z_1 + z_2)   & = 0\\
  (x_0 + x_1 + x_2) & + & (y_1+y_2) & + & (z_0+z_1)     & = 0\\
  (x_1+x_2)         & + & (y_0+y_2) & + & (z_0+z_1+z_2) & = 0
  \end{array}\label{system}\end{equation}

The main idea of the extended binary representation is to represent
the code by a binary graph whose bit-nodes are in one-to-one
correspondence with the set of all possible linear combinations of
$x_i$'s, $y_i$'s, and $z_i$'s. Therefore, we define:
$$S = \left( \begin{array}{*{7}{@{\;}c@{\;}}}
  1 & 0 & 1 & 0 & 1 & 0 & 1 \\
  0 & 1 & 1 & 0 & 0 & 1 & 1 \\
  0 & 0 & 0 & 1 & 1 & 1 & 1
  \end{array}\right)$$
and
$$\begin{array}{rcl}
(\alpha_1,\alpha_2,\dots,\alpha_7) & = & (x_0, x_1, x_2)\times S \\
(\beta_1,\beta_2,\dots,\beta_7) & = & (y_0, y_1, y_2)\times  S \\
(\gamma_1,\gamma_2,\dots,\gamma_7) & = & (z_0, z_1, z_2)\times  S
\end{array}$$
Note that $S$ is the parity check matrix of a Hamming code, thus
$\alpha = (\alpha_1,\alpha_2,\dots,\alpha_7)$, $\beta =
(\beta_1,\beta_2,\dots,\beta_7)$, and $\gamma =
(\gamma_1,\gamma_2,\dots,\gamma_7)$ are codewords of the dual
Hamming code, also called {\em simplex code}. The above linear
equations (\ref{system}) imply that:
$$M_1\alpha + M_2\beta + M_3\gamma = 0,$$
where $M_1$, $M_2$, and $M_3$ are the extended matrices associated
with $h_1$, $h_2$, and $h_3$. This equality corresponds to seven
binary parity checks that can be represented by the binary matrix
below (the zero entries do not appear in the matrix by concern of
legibility). The parity checks $c_1, c_2$, and $c_4$ correspond to
the linear equations of (\ref{system}), and all the other parity
checks ($c_3, c_5, c_6$, and $c_7$) correspond to linear
combinations of the these ones.
$$\begin{array}{c|*{7}{@{\;}c@{\;}}|*{7}{@{\;}c@{\;}}|*{7}{@{\;}c@{\;}}|}
  & \multicolumn{7}{@{\;}c@{\;}}{\alpha} & \multicolumn{7}{@{\;}c@{\;}}{\beta} &  \multicolumn{7}{@{\;}c@{\;}|}{\gamma} \\
  & 1 & 2 & 3 & 4 & 5 & 6 & 7 &
  1 & 2 & 3 & 4 & 5 & 6 & 7 &
  1 & 2 & 3 & 4 & 5 & 6 & 7  \\
  \hline
 c_1 &    &   &   &   & 1 &   &   &     & 1 &   &   &   &   &   &       &   &   &   &   & 1 &  \\
 c_2 &    &   &   &   &   &   & 1 &     &   &   &   &   & 1 &   &       &   & 1 &   &   &   &  \\
 c_3 &    & 1 &   &   &   &   &   &     &   &   & 1 &   &   &   &       &   &   &   & 1 &   &  \\
 c_4 &    &   &   &   &   & 1 &   &     &   &   &   & 1 &   &   &       &   &   &   &   &   & 1\\
 c_5 &    &   & 1 &   &   &   &   &     &   &   &   &   &   & 1 &     1 &   &   &   &   &   &  \\
 c_6 &  1 &   &   &   &   &   &   &     &   & 1 &   &   &   &   &       &   &   & 1 &   &   &  \\
 c_7 &    &   &   & 1 &   &   &   &   1 &   &   &   &   &   &   &       & 1 &   &   &   &   &
   \end{array}$$

\begin{defi}
The matrix $\overline{H}_{\mbox{\scriptsize bin}}$ of size (M(q-1),
N(q-1)), obtained by replacing each coefficient $h$ of $H$ by its
extended binary matrix $M_h$, is called the {\em extended binary
matrix} associated with $H$. The binary code $\overline{\cal
C}_{\mbox{\scriptsize bin}} = \ker(\overline{H}_{\mbox{\scriptsize
bin}})$ is called the {\em extended binary code} associated with
${\cal C}$.
\end{defi}

\begin{defi}
Let $S(p) \in \mbf{M}_{p,q-1}(\gf_2)$ be the binary matrix whose
columns represent the binary decomposition of integers
$j\in\{1,\dots,q-1\}$. The {\em simplex code} ${\cal S}(p)$ is the
$[q-1, p, 2^{p-1}]$ linear binary code with generator matrix $S(p)$.
\end{defi}

\begin{theo} Let $\overline{\cal
C}_{\mbox{\scriptsize bin}}$ be the extended binary code associated
with a non binary code ${\cal C}$.

\noindent  (1) Let $(X_1,\dots,X_N)\in{\cal C}$, and for each $ n
  \in\{1,\dots,N\}$ let
  $(\alpha_{n,1}, \dots,
  \alpha_{n,q-1}) \in {\cal S}(p)$ be the simplex codeword
  obtained by encoding the binary image $(x_{n,0}, \dots, x_{n,p-1})$ of $X_n$.
  Then $$(\alpha_{1,1}, \dots,
  \alpha_{1,q-1},\dots\dots,\alpha_{N,1}, \dots,
  \alpha_{N,q-1})\in\overline{\cal C}_{\mbox{\scriptsize bin}}$$

\noindent  (2) The above mapping defines a vector space isomorphism:
$${\cal C}\stackrel{\sim}{\longrightarrow} \overline{\cal C}_{\mbox{\scriptsize bin}}
\cap {\cal S}(p)^N$$
 where ${\cal S}(p)^N = {\cal S}(p) \times\cdots\times {\cal S}(p) \subset
 \gf_2^{N(q-1)}$ is the vector space product of $N$ copies of
 ${\cal S}(p)$.
%
\end{theo}

An intuitive interpretation of the above theorem is that a
non-binary code can be represented by a graph with $N(q-1)$
bit-nodes and $M(q-1)$ check-nodes connected according to the
extended binary matrix $\overline{H}_{\mbox{\scriptsize bin}}$, and
$N$ simplex-nodes connected each one to $(q-1)$ consecutive
bit-nodes. Hence, within a message-passing decoding, the bit-nodes
should recover their values from messages received from both simplex
and check-nodes of the graph. Although we are interested in decoding
non-binary codes over the BEC, the ideas presented in this paper
might be extrapolated to other channels.

\begin{rema}
The extended binary representation is also useful for understanding
aspects related to cycles of the bipartite graph associated with a
 non-binary LDPC code. Assume that all the non-zero entries of $H$ are
invertible. Let $\overline{\cal H}_{\mbox{\scriptsize bin}}$ be the
bipartite graph associated with the matrix
$\overline{H}_{\mbox{\scriptsize bin}}$. It follows from the
construction that $\overline{\cal H}_{\mbox{\scriptsize
 bin}}$ is a covering graph of ${\cal H}$, hence any cycle of $\overline{\cal
H}_{\mbox{\scriptsize bin}}$  {\em lies over} some cycle of ${\cal
H}$. Furthermore, let
 $(e_1,e_2,\dots, e_{2\ell})$ be a cycle of length
 $2\ell$ of ${\cal H}$, and let $h_i$ denote the label of the edge $e_i$. Then, the number and the length of cycles
 of $\overline{\cal H}_{\mbox{\scriptsize
 bin}}$ lying over $(e_1,e_2,\dots, e_{2\ell})$ can be derived
 using the cycle decomposition of the permutation $\Phi_h$, where
 $h = h_1h_2^{-1}\cdots h_{2\ell-1}h_{2\ell}^{-1}$, in a similar way
 as for
 quasi-cyclic codes (see for instance \cite{fossorier2004qld}).
\end{rema}

\section{Linear time erasure decoding}
\label{linear_time_decoding} Similar to section \ref{gf_dec_bec}, we
assume that a non-binary LDPC code is used over the BEC($\epsilon$).
Let $(X_1, X_2,\dots,X_N)$ be the length-$N$ sequence of encoded
symbols, and let $(x_{1,0},\dots,x_{1,p-1}, \dots\dots
x_{N,0},\dots,x_{N,p-1})$ denote its binary image of length $Np$,
which is transmitted over the BEC; each of its bits being erased
with probability $\epsilon$. At the receiver part, the received bits
are used to provide information to the corresponding bit-nodes in
the extended binary graph $\overline{\cal H}_{\mbox{\scriptsize
bin}}$. More precisely, for each coded symbol $X_n$ there are $q-1$
corresponding bit-nodes in $\overline{\cal H}_{\mbox{\scriptsize
bin}}$, which are denoted by $(\alpha_{n,1}, \alpha_{n,2},\dots,
\alpha_{n,q-1})$. Recall that each $\alpha_{n,k}$ corresponds to a
linear combination of $x_{n,0},\dots,x_{n,p-1}$, whose coefficients
are given by the binary decomposition of $k\in\{1,\dots,q-1\}$.
Therefore, the bit-node $\alpha_{n,2^i}$, $0\leq i\leq p-1$,
corresponds to the bit $x_{n,i}$ from the binary sequence that is
transmitted over the BEC.

The decoding algorithm is initialized as follows:
\begin{itemize}
  \item for each received bit $x_{n,i}$ set:

  \vspace{-3mm}
  $$\alpha_{n,2^i} = x_{n,i}$$
  \item set all the other bit-nodes $\alpha_{n,k}$ as erased
\end{itemize}
Note that a bit-node $\alpha_{n,k}$ is set at erased if either $k$
is not a power of $2$, or $k = 2^i$ but the corresponding bit
$x_{n,i}$ was erased by the channel. Erased bit-nodes are then
iteratively recovered as follows:
\begin{itemize}
  \item {\bf simplex-node processing}

  for each $n\in\{1,\dots,N\}$, if bit-nodes
  $\alpha_{n,k_1},\dots,\alpha_{n,k_i}$ are recovered (either received or recovered at the previous iterations),
  recover the value of $\alpha_{n,k_1\wedge\cdots\wedge k_i}$ by:
  $$\alpha_{n,k_1\wedge\cdots\wedge k_i} =\alpha_{n,k_1}\wedge\cdots\wedge\alpha_{n,k_i}$$
  \item {\bf check-node processing}

  for any check-node $c\in \overline{\cal H}_{\mbox{\scriptsize bin}}$ connected to a single unrecovered bit-node
  $\alpha_{n,k}$, recover
  the value of $\alpha_{n,k}$ as the XOR of the other bit-nodes
  connected to $c$.
\end{itemize}
The simplex-node processing and the check-node processing are
iterated as long as new bit-nodes $\alpha_{n,k}$ can be recovered.
The decoding is successful if all the bit-nodes are recovered when
it stops.

It is important to note that the above decoding is equivalent to the
non-binary decoding presented in section \ref{gf_dec_bec}. There is
a one-to-one correspondence between recovered bit-nodes and sets of
eligible symbols, which can be described as follows:
\begin{itemize}
  \item Let $R_n$ be the set of all recovered bit-nodes
  $\alpha_{n,k}$ after the simplex-node
  processing step, for some $n\in\{1,\dots,N\}$. Each
  $\alpha_{n,k}\in R_n$ gives the value of some linear combination
  of bits $x_{n,0},\dots, x_{n,p-1}$, that is:
  $$\sum_{i=0}^{p-1} k_ix_{n,i} = \alpha_{n,k},$$
  where $(k_0, \dots, k_{p-1})$ is the binary decomposition of $k$.
  Let ${\cal E}_n \subset {\cal A}$ be the subset of all symbols
  whose binary images verify the above equation for all $\alpha_{n,k}\in
  R_n$. Then ${\cal E}_n$ coincides with the affine subspace of eligible symbols
  defined in section \ref{gf_dec_bec}. The fact that ${\cal E}_n$ is
  affine follows from the fact that for any $\alpha_{n,k}, \alpha_{n,l} \in
  R_n$, we also have $\alpha_{n,k\wedge l}\in R_n$.

  \item Let $m$ be a constraint-node of the non-binary graph ${\cal
  H}$ and let $c_1, \dots, c_{q-1}$ be the corresponding parity
  check-nodes in the binary graph $\overline{\cal H}_{\mbox{\scriptsize
  bin}}$. Let $n\in{\cal H}(m)$, and denote by $R_{m,n}$ the set of all bit-nodes $\alpha_{n,k}$
  that are recovered by the check-nodes $c_1, \dots, c_{q-1}$ from the
  unerased nodes among the bit-nodes $\alpha_{n',k'}$, with $n'\in{\cal
  H}(m)\setminus\{n\}$. By using the same arguments as above, $R_{m,n}$ defines a subset ${\cal E}_{m,n} \subset {\cal
  A}$, which coincides with the affine subspace of eligible symbols
  defined in section \ref{gf_dec_bec}. The fact that ${\cal E}_{m,n}$ is
  affine follows from the fact that whenever $\alpha_{n,k}$ and
  $\alpha_{n,l}$ are recovered by check-nodes $c_i$ and $c_j$,
  then
  $\alpha_{n,k\wedge l}$ is also recovered by the check-node $c_{i\wedge
  j}$.
\end{itemize}

We discuss now the complexity of the proposed erasure decoding. The
processing of each check-node is done in constant time. Since the
number of check-nodes in $\overline{\cal H}_{\mbox{\scriptsize
  bin}}$ depends linearly on $q$, it follows that the check-node processing step of
  the decoding algorithm is done in linear time. Moreover,
   the simplex-node processing can also be implemented in linear time.
   Fix some $n\in\{1,\dots,N\}$, and let $R_{\mbox{\scriptsize
  in}}$ and $R_{\mbox{\scriptsize out}}$ denote the sets of
  recovered bit-nodes $\alpha_{n,k}$ before and after the
  simplex-node processing. Then $R_{\mbox{\scriptsize out}}$ is the
  ``affine subspace'' spanned by $R_{\mbox{\scriptsize in}}$, in the sense that
  $R_{\mbox{\scriptsize out}} \supseteq R_{\mbox{\scriptsize in}}$
  and $\alpha_{n,k\wedge l}\in R_{\mbox{\scriptsize out}}$ for any
  $\alpha_{n,k}, \alpha_{n,l}\in R_{\mbox{\scriptsize out}}$,
  and it can
  be computed as follows:

  \smallskip
  \noindent $R_{\mbox{\scriptsize out}} = \{\}$\\
  while $R_{\mbox{\scriptsize in}}$ is not empty\\
  $\mbox{}$\hspace{5mm}  $\alpha_{n,k} \leftarrow R_{\mbox{\scriptsize in}}.{\mbox{pop}}()$\\
  $\mbox{}$\hspace{5mm}  $R_{\mbox{\scriptsize tmp}} = R_{\mbox{\scriptsize out}}$ \\
  $\mbox{}$\hspace{5mm}  for $\alpha_{n,l} \in R_{\mbox{\scriptsize tmp}}$\\
  $\mbox{}$\hspace{10mm} $\alpha_{n,k\wedge l} = \alpha_{n,k}\wedge\alpha_{n,l}$\\
  $\mbox{}$\hspace{10mm} $R_{\mbox{\scriptsize out}} = R_{\mbox{\scriptsize out}} \cup \{\alpha_{n,k\wedge l}\}$\\
  $\mbox{}$\hspace{10mm} $R_{\mbox{\scriptsize in}} \ = R_{\mbox{\scriptsize in}}\ \setminus \,\{\alpha_{n,k\wedge l}\}$\\
  $\mbox{}$\hspace{5mm}  end\\
  $\mbox{}$\hspace{5mm} $R_{\mbox{\scriptsize out}} = R_{\mbox{\scriptsize out}} \cup \{\alpha_{n,k}\}$\\
  end

It can be easily seen that the above implementation requires
$1+2^1+\cdots + 2^{\mid R_{\mbox{\scriptsize in}}\mid -1} \leq 2^p-1
= q-1$ computations, where $|R_{\mbox{\scriptsize in}}|$ denotes the
{\em dimension} of the vector subspace of $\mathbb{Z}_q$ spanned by
$\{k\mid \alpha_{n,k}\in R_{\mbox{\scriptsize in}}\}$.

The above discussion is resumed by the following:
\begin{theo}
  The complexity of the extended binary erasure decoding of non-binary LDPC codes depends
  linearly on the size of the alphabet.
\end{theo}

Before concluding the paper, we would like to emphasis some other
advantages of the extended binary decoding. These aspects will be
developed in future works.

1) {\em Stopping sets}. Similar to binary LDPC codes, we can define
{\em stoping sets}, corresponding to erasure patterns from which the
decoding cannot recover. Thus, a stopping set is a subset
$\mathscr{S}$ of the set of bit-nodes of $\overline{\cal
H}_{\mbox{\scriptsize bin}}$, such that:
\begin{itemize}
  \item if $\alpha_{n,k\wedge l} \in \mathscr{S}$ then either $\alpha_{n,k} \in \mathscr{S}$ or $\alpha_{n,l} \in \mathscr{S}$
  \item check-nodes that are neighbors of $\mathscr{S}$ are connected to
  $\mathscr{S}$ at least twice.
\end{itemize}
Hence, the finite length analysis of non-binary LDPC codes over the
BEC can be derived by using techniques similar to those developed in
\cite{di2002fla}.

 2) {\em UL-FEC applications}. In practical systems, data packets
 received at the upper-layers encounter erasures, and
 erasure codes are used to recover the erased data packets.
  If non-binary LDPC codes are used in such situations,
 the coded symbols must be {\em transverse} to data packets: that is, the $p$ bits of a symbol
 must belong to $p$ different data packets (otherwise if, for instance, all the $p$ bits of a symbol
 belong to the same data packet, the coded symbols will be either completely received or completely erased,
 and the non-binary code would operate as a binary code)\footnote{This is contrasting with
 other non-binary UL-FEC codes, as the Reed-Solomon codes, for which the $p$ bits of a
 symbol must belong to the same data packet.}. The ability of the decoding algorithm of dealing with
 data packets instead of dealing with bits is an attractive feature of an erasure
 code. The proposed extended binary decoding is well-suited for UL-FEC applications as it
 can easily deal with data packets: the bit-nodes $\alpha_{n,k}$ would
 correspond to packets instead of a single bit, but the decoding
 would work the same way, simply by performing bitwise XOR of
 packets $\alpha_{n,k}$.

\begin{figure}[!t]
\noindent\includegraphics[width=\linewidth]{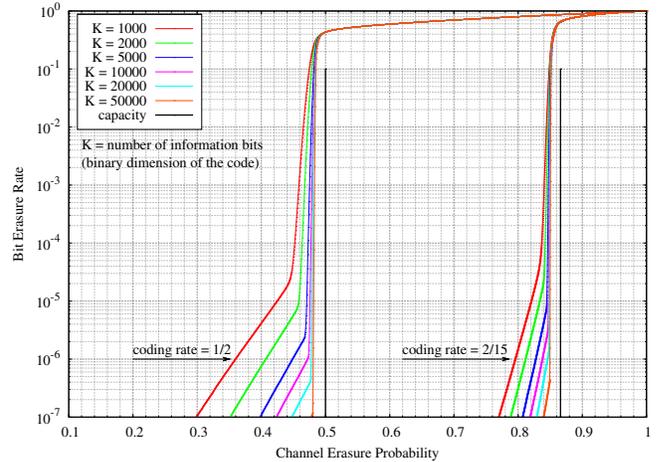}

\vspace{-3mm}
 \caption{Incremental redundancy using non-binary LDPC
codes} \label{incremental_redundancy} \vspace{-5mm}
\end{figure}

3) {\em Flexibility and small coding rates}. Another interesting
feature of the proposed decoding is the possibility of using
incremental redundancy in order to cope with severe channel
conditions. This can be done by transmitting all the
 $N(q-1)$ values of the bit-nodes $\alpha_{n,k}$ over the channel,
instead of transmitting only the $Np$ bits $x_{n,i}$ of the binary
image. This is illustrated in Figure \ref{incremental_redundancy}.
We use an irregular LDPC code over $\gf_{16}$, with rate $r = 1/2$.
In case that all the $N(q-1)$ values of the bit-nodes $\alpha_{n,k}$
are transmitted over the channel, the coding rate is decreased to
$r' = \displaystyle r\frac{p}{q-1} = 2/15$. As it can be seen, in
both situations, the code operates very close to the channel
capacity. For large values of $q$, the incremental redundancy turns
the code into an {\em almost rateless} code.

\section{Conclusions}
\label{conclusions}

We showed that non-binary LDPC codes can be described in terms of
binary parity-check and simplex constraints. On the one hand, this
description can be used for decoding non-binary LDPC codes, and the
proposed decoding presents several attractive properties for
practical applications: low complexity, capability of dealing with
data packets for UL-FEC applications, on-the-fly decoding,
incremental redundancy, and small coding rates. On the other hand,
the proposed description gives insights into the structure of
non-binary codes, and is very likely that it might be used for both
finite length and asymptotical analysis of non-binary LDPC codes.

 \bibliographystyle{./bib/IEEEbib}
\footnotesize
\bibliography{./bib/MyBiblio,./bib/Zotero}

\end{document}